\documentclass[review]{elsarticle}

\usepackage{algorithm}
\usepackage[noend]{algpseudocode}
\usepackage{amsmath}
\usepackage{hyperref}

\journal{Journal of \LaTeX\ Templates}

\begin{document}

\begin{frontmatter}

\title{Vericom: A Verification and Communication Architecture for    IoT-based Blockchain}


\author[1]{Ali Dorri}

\author[2]{Shailesh Mishra} 

\author[1] {Raja Jurdak}

\address[1]{School of Computer Science, QUT, Australia.}
\address[2]{Department of Electrical Engineering, IIT Kharagpur, India}

\begin{abstract}
Blockchain has received tremendous attention  as a secure, distributed, and  anonymous framework for the Internet of Things (IoT).  As a distributed system, blockchain trades off scalability for distribution, which limits the technology’s adaptation for large scale networks such as IoT.   All transactions and blocks must be  broadcast  and verified by all participants which limits scalability and incurs computational and communication overheads. The existing solutions  to scale blockchains have so far led to partial recentralization, limiting the technology’s original appeal.  In this paper, we introduce   a distributed yet scalable  Verification and Communication architecture for blockchain referred to as Vericom. Vericom concurrently achieves high scalability and distribution using hash function outputs to shift blockchains from broadcast  to multicast communication. Unlike conventional blockchains where all nodes must verify new transactions/blocks, Vericom  uses the hash of IoT traffic to randomly select a set of nodes to verify transactions/blocks which in turn reduces the processing overhead. 	  Vericom incorporates two layers: i) transmission layer where a randomized multicasting method is introduced along with a backbone network to route traffic, i.e., transactions and blocks, from the source to the  destination, and ii) verification layer where a set of randomly selected nodes are allocated to verify each transaction or block.  The performance evaluation shows that Vericom reduces the packet and processing overhead as compared with conventional blockchains. In the worst case,   packet overhead in Vericom scales linearly with the number of nodes  while  the  processing overhead remains scale-independent.
\end{abstract}


\end{frontmatter}


\section{Introduction}\label{sec:intro}
The Internet of Things (IoT) is a network of millions of low-resource devices that collect and exchange information about the physical environments  which is then processed by  service providers (SPs) to offer personalized services to the users.  Conventional IoT ecosystems rely on a brokered communication model where the communications, authentication, and authorizations are conducted by a  central trusted authority. In many situations, geographically proximate IoT devices still have to go through a remote central server  to access services which is  unlikely to scale when millions of nodes are connected. SPs collect a huge volume of personalized information about the users and thus can build a virtual profile about them which risks user  privacy. The conventional security architectures are not directly applicable in IoT as IoT encompasses heterogeneous low-resource devices which come with no or limited built-in security   features \cite{zhang2014iot,khan2018iot}.  Most of the existing  IoT-specific security solutions largely rely on centralized communication models which suffer from lack of scalability and single point of failure \cite{hassan2019current,alaba2017internet}.  \par

In recent years,  blockchain has received tremendous attention to address the outlined challenges in IoT due to  its salient features including decentralization, anonymity, trust,  and  security \cite{dai2019blockchain}.   Blockchain is an immutable    database shared across all participating nodes in the network and was first introduced in Bitcoin \cite{nakamoto2019bitcoin}, the first cryptocurrency, in 2008. A transaction represents the basic communication primitive between the participating nodes which is sealed using asymmetric encryption.  Blockchain participants  are known by a unique  PK that can be changed for each transaction which in turn  introduces a  level of anonymity.  All transactions are broadcast in the network and verified by all participants. Transaction verification typically involves matching the Public Key (PK) with the associated signature (both are stored in the transaction). Some nodes, known as validators, may choose to  store new transactions in blockchain in the form of a block which requires  following a consensus algorithm. The latter protects  blockchain  security against malicious validators that may attempt to flood the network with  fake blocks and ensures the validator of the next block is selected randomly. This introduces a trusted network where untrusted  participants can exchange information. \par

\begin{figure}
	\begin{center}
		\includegraphics[width=8cm ,height=7cm ,keepaspectratio]{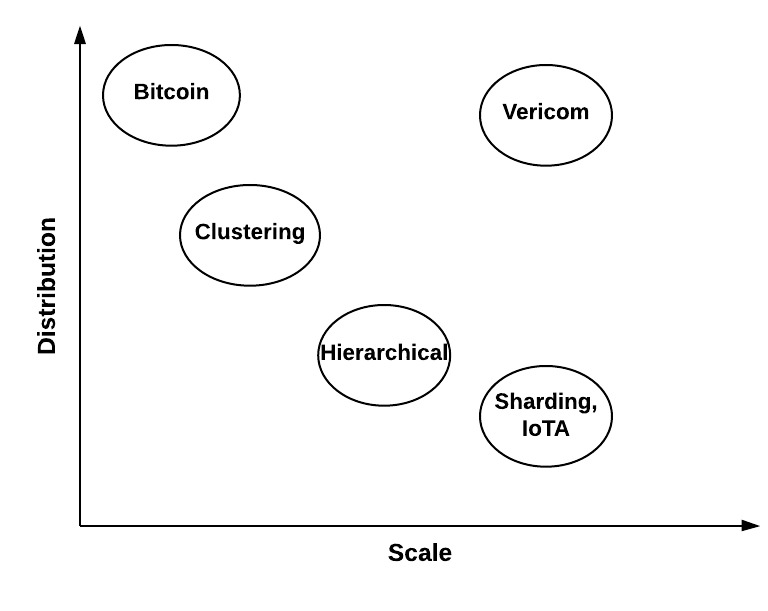}   
		\caption{The traditional correlation between distribution and scale for IoT-based blockchain. }
		\label{fig:dis-scale-diagram}
	\end{center}
\end{figure}

Despite their significant advantages,  conventional blockchains are not directly applicable in IoT due to the lack of scalability resulting from huge communication and computational overheads. New transactions and blocks are broadcast to all IoT nodes, i.e., IoT devices and users,  to ensure distributed management of the blockchain. This  in turn demands  significant bandwidth which is far beyond limited energy, processing, and communication capabilities  of the IoT nodes. IoT nodes have limited energy resources. Packet transmission is among the most energy consuming  tasks in IoT nodes which  makes it impossible for them  to participate in the management of the blockchain, where the participants are expected to receive a huge volume of transactions/blocks. New transactions/blocks must be verified by all nodes which in turn requires significant computational resources that is  far beyond the capabilities of IoT nodes. \par  

\begin{figure*}
	\begin{center}
		\includegraphics[width=14cm ,height=13cm ,keepaspectratio]{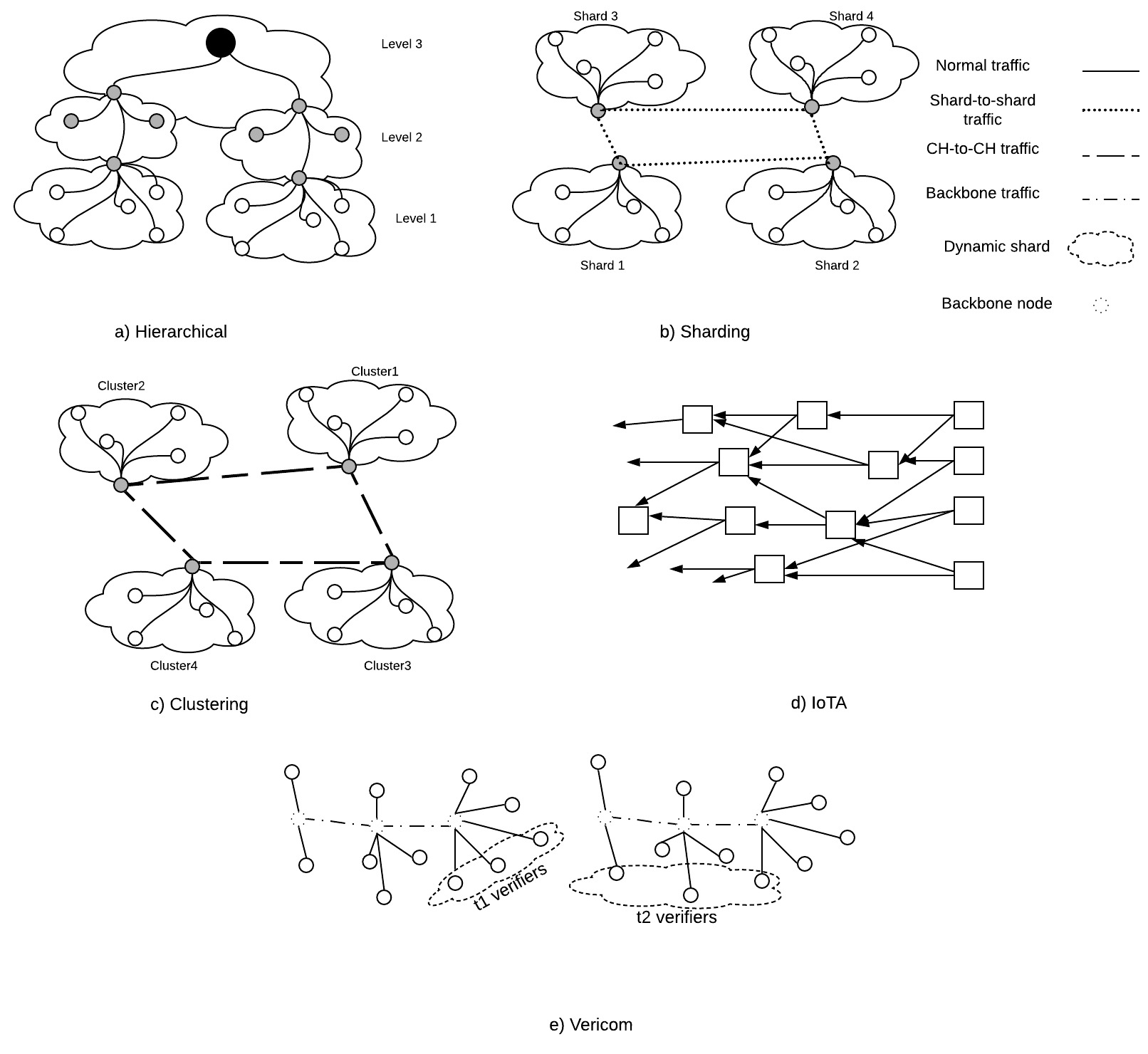}   
		\caption{A high-level topological view of the studied categorizes. (a),(b),(c) create static groups with large number of participants to limit the nodes in the blockchain  while (d) represents IoTA where transactions are linked  (e)  forms dynamic group with small participants that verify transactions.}
		\label{fig:architectures}
	\end{center}
\end{figure*}

There is a known trade-off  between scalability and distribution as shown in Figure \ref{fig:dis-scale-diagram}. Purely distributed chains, such as Bitcoin \cite{nakamoto2019bitcoin}, suffer from lack of scalability due to their reliance on broadcast communication and computationally demanding block and transaction verification processes where  all transactions and blocks needs to be verified by all the participating nodes.  Existing proposals to enhance blockchain scalability include heirarchical, sharded, or clustered blockchains (see Figure  \ref{fig:architectures})   \cite{ma2019privacy,tong2019hierarchical,dorri2019lsb}. In hierarchical methods \cite{dorri2019lsb},  multiple levels of hierarchy are created where the transactions and blocks in each hierarchy are only broadcast to the nodes at the same level in the heirarchy  (see Figure \ref{fig:architectures}.a).  In sharding \cite{yun2020dqn}, the network is divided into multiple groups, i.e., shards, where the transactions of each shard are only broadcast to the nodes in the same shard (see Figure \ref{fig:architectures}.b). In clustering algorithms, the network is grouped into multiple clusters where a high resource available node, known as cluster head (CH),  forwards blocks and transactions to/from the cluster members (see Figure \ref{fig:architectures}.c). Unlike sharding where each shard functions independently, in clustering the  transactions and blocks are broadcast and the CHs  jointly manage the blockchain. In addition to the conventional blockchains, IoTA, a distributed ledger technology, has been introduced to eliminate centralization in IoT. However, IoTA still relies on broadcast communications and suffers from centralization of a coordinator node that verifies transactions.

As shown in Figure \ref{fig:dis-scale-diagram} the existing methods sacrifice distribution for  scale by deviating from blockchain's original distributed topology. The security and anonymity of the blockchains are directly impacted by the distribution and scaling   features. A distributed large scale network  achieves higher security as compared with more centralized smaller scale chains as a larger number of potential validators exist in the network, making it harder  for  attackers to  dominate the blockchain network and store fake blocks. Additionally, blocks are verified by a broader set of nodes which protects  against colluding nodes that may mark a fake block as valid.  From the anonymity perspective, the large number of participants continuously increases the number of transactions and PKs   in the blockchain  which in turn complicates user deanonymization \cite{dorri2019activity} by linking a group of transactions or PKs to a particular IoT node.     Evident from the above discussions, it is critical to reduce the blockchain communication and computational overheads without  sacrificing the distributed feature.  \par

The main contribution of this paper is to push the boundaries of  distribution and scale  (see Figure \ref{fig:dis-scale-diagram}) to achieve a distributed yet scalable blockchain which is adoptable in IoT ecosystem by introducing a   verification  and communication architecture known as  \textit{Vericom}.  Vericom incorporates two layers  namely: \par 

i) Transmission layer: In this layer,  we introduce a multicasting algorithm that sends traffic, i.e., transactions and blocks, to a group of IoT nodes that are selected randomly,  dynamically, and  in an unpredictable manner based on the hash of the traffic. This shift from broadcast (as in conventional blockchains) to multicast  significantly reduces  the bandwidth consumption of the underlying IoT nodes while the randomness, uniqueness, and dynamic allocation  of the destination group avoids  centralization.  Vericom incorporates a backbone network  that receives traffic from the source IoT node  and employs IP-based routing algorithms to route the traffic toward the destination IoT nodes   that are randomly selected based on the hash of the transaction/block (details in  verification layer).  The backbone network is similar to the Internet backbone network that routes the Internet traffic even in the existing blockchain architectures. To incentivize nodes to function as backbone nodes, Vericom introduces a traffic management fee (TMF) that is paid by the validators to the backbone nodes. The value of TMF  is defined based on the    number of blocks generated by a validator  during an epoch time.    \par

ii) Verification layer: In this layer  a subset of the participating nodes, known as verifier set, are randomly and dynamically dedicated to  verify a particular transaction or block which in turn reduces the computational overhead associated with transaction/block verification on the IoT devices  as compared with conventional blockchains where all transactions and blocks are broadcast. The main aim of this layer is to limit the verifiers of the traffic while preserving the security of the ledger against malicious nodes that may mark an invalid block or transaction as valid  to store fake data in the blockchain.   The selection of the verifier set  is based on the hash of the transaction/block content that ensures randomness and unpredictability of the verifier set. This also ensures the distributed nature of Vericom as for each transaction/block a unique verifier set is randomly and dynamically identified. Once a block is verified by the verifier set, it is broadcast to the network and any other node may also attempt to verify the same to protect against rare cases where all the nodes in the verifier set are colluding malicious nodes. \par 

We study the security of Vericom against three possible  attacks and discuss how Vericom is resilient against such attacks.   The simulation results show that  Vericom reduces the packet and computational overhead as compared with conventional blockchains. \par 

The rest of the paper is organized as follows. Section \ref{sec:lit-review} studies  the existing solutions to enhance the blockchain scalability.  
Section \ref{sec:architecture} outlines the details of Vericom.  Section \ref{sec:evaluation} presents the evaluation results and discusses the performance of Vericom and finally Section \ref{sec:conclusion} concludes the paper and outlines future research directions. 

\section{On the Scalability of Blockchain} \label{sec:lit-review}
In this section, we study the existing solutions in the literature to enhance the blockchain scalability for IoT.   Let us first explain the blockchain scalability limitation in a  smart grid setting, as a representative use case  of IoT,  to motivate the discussions in the rest of this section. The penetration of Distributed Energy Resources (DER)   and  smart meters leads to the emergence of energy prosumers which are the nodes capable of consuming and producing energy. A smart home equipped with solar panels is an example of an energy prosumer. The prosumers may trade their excess energy with other energy consumers which leads to a significant volume of transactions as trading energy involves broadcasting multiple transactions to find consumers/producers and to negotiate trade terms with the intended seller/buyer  \cite{khorasany2021lightweight}.  Additionally, the grid participants  may need to frequently (in 1 hour or 0.5 hour intervals) send  data to the grid operator, or other trusted parties,  to ensure the reliability of the grid and balance the energy demand with supply. The transactions and blocks need to be verified by the blockchain participants which in turn demands significant computational resources from the participants while smart grid participants, e.g. solar panels and  smart meters, have limited resources.\par 

Having  motivated the need for blockchain scalability in cyberphysical environments,  we next study the existing optimization methods  in the literature to enhance blockchain scalability. Based on the communication topology employed by each optimization method, we classify those in three categories namely hierarchical methods, sharding, and clustering. A high-level topological view of these categories is shown in Figure \ref{fig:architectures}.

\subsection{Hierarchical Methods}
In hierarchical methods, as shown in Figure \ref{fig:architectures}.a, the network is  organized in the form of multiple hierarchies where the transactions in each level of the hierarchy are only broadcast to the nodes in the same level which in turn enhances the blockchain scalability. In each level, a manager node functions as the central authority that authorizes nodes to join the network at that level and forwards traffic to/from the upper level hierarchies. The authors in \cite{ma2019privacy} proposed a hierarchical architecture for access control management in blockchain. The framework comprises  three main layers which are device, fog, and cloud layers. The device layer  encompasses the IoT devices. The fog layer includes the first tier of the blockchain that is managed by a central node. Each device in the device layer  associates with a node in the fog layer that stores the transaction of the device in a private chain. The cloud layer comprises  central servers that run the global blockchain and  store a copy of the private blockchains. \par 

The authors in \cite{oktian2020hierarchical} proposed a hierarchical architecture that comprises  four chains namely: 1) payment engine that handles the payments and micropayments, 2) compute engine that governs the smart contracts and runs distributed applications, 3) storage engine that stores the data associated with transactions, and 4) core engine that comprises of  a public ledger that connects all chains and enables transaction exchange between other chains, i.e., engines. \par  

\textit{Side chains} are introduced to enhance the blockchain scalability and reduce delay in trading assets \cite{back2014enabling,singh2020sidechain}. To create a side chain, a user locks a particular amount of asset  in the main chain and transfers it to the side chain where it can be traded with other parties without requiring the transactions to be sent to the main chain. Once the trade is concluded, the asset ownership information is transferred to the parent chain and the side chain is closed \cite{singh2020sidechain}.

\subsection{Sharding}
Sharding refers to partitioning the network where each partition, also known as shard, functions  independently  and is managed by a shard manager (see Figure \ref{fig:architectures}.b). The transactions and blocks generated by the  nodes in each shard are only broadcast and verified  in the same shard which in turn increases  scalability. Shard-to-shard communication is limited to where the verification of  a transaction in one shard requires input from a transaction in another shard,   e.g., spending the output of a transaction in a different shard. In such cases, the managers of the involved shards communicate to verify the transaction. \par 

The authors in \cite{tong2019hierarchical} proposed a sharded blockchain architecture to enhance scalability of the blockchain for IoT applications.  In each shard  a manager node  authorizes the nodes that can join the shard,  verifies new transactions, and stores  new blocks.  The shards are connected through a main shard  where the shard managers  connect to reach  consensus on sub-blocks, i.e., the chains in each shard.  

\subsection{Clustering}
Clustering   enhances  blockchain scalability for IoT by reducing the number of nodes that participate in blockchain management. The network is clustered into multiple groups  (see Figure \ref{fig:architectures}.c). In each cluster, a node with sufficient computational resources  is selected as the cluster head (CH) that i)  receives transactions from the cluster members and broadcasts to the blockchain, ii) participates in the blockchain by verifying new transactions and storing blocks, and iii) forwards transactions to the cluster members if they are the destination. \par 

Unlike sharding where communication between shard managers is limited to selected transactions,   clustering broadcasts all blocks and transactions  between the CHs.  In \cite{dorri2019lsb} we introduced a scalable blockchain where new blocks and transactions are broadcast and verified only by the CHs. The cluster members populate an  Access Control List (ACL) to authorize particular nodes in the network to send them transactions. CHs employ the ACL to decide whether to send a transaction to the cluster members or to other CHs.

\subsection{IoTA}
IoTA \cite{popov2018tangle}  has been introduced as a scalable solution that can address the limitations of blockchain for large scale IoT. IoTA is not a blockchain but a distributed ledger technology, where unlike blockchains transactions are not committed in the form of blocks. Instead, IoTA introduced the \textit{Tangle} (see Figure \ref{fig:architectures}.d) that creates a directed acyclic graph (DAG) to store transactions. In order for a transaction to be stored in IoTA, the transaction generator must randomly select and verify two previously generated transactions. As more nodes verify a transaction, and thus more transactions are chained to it, the weight and thus confirmation level of the transaction increases.

\subsection{Discussion}
Having discussed the key methods employed to increase blockchain scalability, we next evaluate such methods and highlight their limitations for IoT. \par 

In the outlined methods, the network is divided into subgroups which are managed by a central manager which in turn leads to  a  decentralized topology. The manager  still suffers from the centralization challenges including privacy, security, single point of failure and scale.  The reduced number of participants in each group, facilitates user deanonymization as  the malicious nodes  deanonymize the users from a smaller pool.  In the above methods, there is a trade-off between the number of groups, e.g., clusters, and the centralization degree. Fewer groups increase scalability but lead to more centralization.  \par 

In hierarchical methods, the packets will eventually traverse multiple tiers before being considered as valid which in turn increases the delay in verifying transactions.  In sharding methods,  intra-shard communication and verification remain challenging and incur significant delay.  \par 

IoTA relies on broadcast communications which in turn suffers from high packet overhead as in conventional blockchains. IoTA improves the verification processing overhead as fewer nodes confirm a transaction, however, it comes with the cost of increased delay as transactions shall wait for longer time to receive enough weight. Additionally, IoTA relies on a coordinator node that centrally confirms transactions each two minutes which in turn suffers from centralization and moves away from distributed technology \cite{silvano2020iota}. 

Vericom introduces a distributed yet scalable blockchain by multicasting traffic to a randomly selected set of nodes that verify a transaction/block (see Figure \ref{fig:architectures}.e). The verifier set is unique for each transaction/block and is selected randomly and dynamically based on the hash of the transaction/block content which in turn increases blockchain security against malicious nodes that may store fake transactions. The verifier set dynamically changes per transaction/block which moves away from centralization and thus mitigates the related challenges such as privacy and security. Any of the participating nodes in  blockchain may also choose to verify new blocks or transactions to detect misbehavior. Malicious nodes are isolated from the network to mitigate the impact of attacks. By introducing a distributed yet scalable architecture,  Vericom support a larger number of participating nodes   which in turn reduces the chance of user deanonymization as  a huge volume of transactions are stored in the main chain.\par 
Depending on the read/write  permissions of the participating  nodes, blockchain can be categorized as \cite{wang2019survey}: i) \textit{Public} where all nodes have equal read/write permissions and any node can participate on blockchain, and ii) \textit{Private} where authorized nodes may only have write permission, i.e., store new blocks and a selected node authorizes the nodes that can participate on the network. The discussion in the rest of this paper applies to both public and private chains, however, the maximum benefit is for the public chains due to the large scale and openness of such ledgers. \par

\section{Vericom: A Verification and Communication Architecture for IoT-based Blockchain}\label{sec:architecture}

This section outlines the details of Vericom by discussing the preliminaries in Section \ref{sec:background} and the core Vericom algorithms in Section \ref{sec:vericom-core}.

\subsection{Preliminaries} \label{sec:background}
Vericom delivers a distributed yet scalable blockchain architecture for IoT by significantly reducing the communication overhead, achieved through dynamic multicast, and the computational overhead, achieved through the use of hash function outputs for randomised and secure verifer set selection. Vericom is part of a bigger project with the aim of designing an IoT-friendly blockchain architecture. In our earlier work, Tree-chain \cite{dorri2020tree}, we introduced a fast consensus algorithm where a leader is selected randomly to commit transactions that match a particular pattern in the blockchain. Tree-chain is a lightweight validator selection algorithm (also known as consensus algorithm in the literature) that significantly reduces  the computational overhead and delay associated with storing new blocks by randomizing the selection of validator sets  based on  hash function output.   The upper-bound throughput of the Tree-chain is the speed at which a validator can verify a transaction which is in near real-time. Such fast transaction commitment requires fast transaction delivery to the validators which is impossible in the existing blockchains due to the broadcast nature. Each validator has limited bandwidth which limits the number of transactions/blocks it can receive and thus impacts the blockchain throughput. In other words, Tree-chain shifts the transaction   throughput  bottleneck  from the consensus algorithm to the packet propagation and transaction delivery. Vericom aims to address this limitation. In this paper and without loss of generality, we assume that Tree-chain  is employed as the underlying validator selection algorithm. However, Vericom is applicable with any other validator selection algorithm.  \par 

 \par

 In Tree-chain the randomization among validators is achieved in two levels:\par 
i) Transaction level where the validator of each transaction is selected randomly based on the hash of the  transaction content. Each validator commits transactions  whose hash value  starts with a particular  character, known as the consensus code.

ii) Block level where the consensus code corresponding to each validator is randomly allocated  based on the hash of the PK of the validator. \par 

\begin{figure}
	\begin{center}
		\includegraphics[width=7cm ,height=7cm ,keepaspectratio]{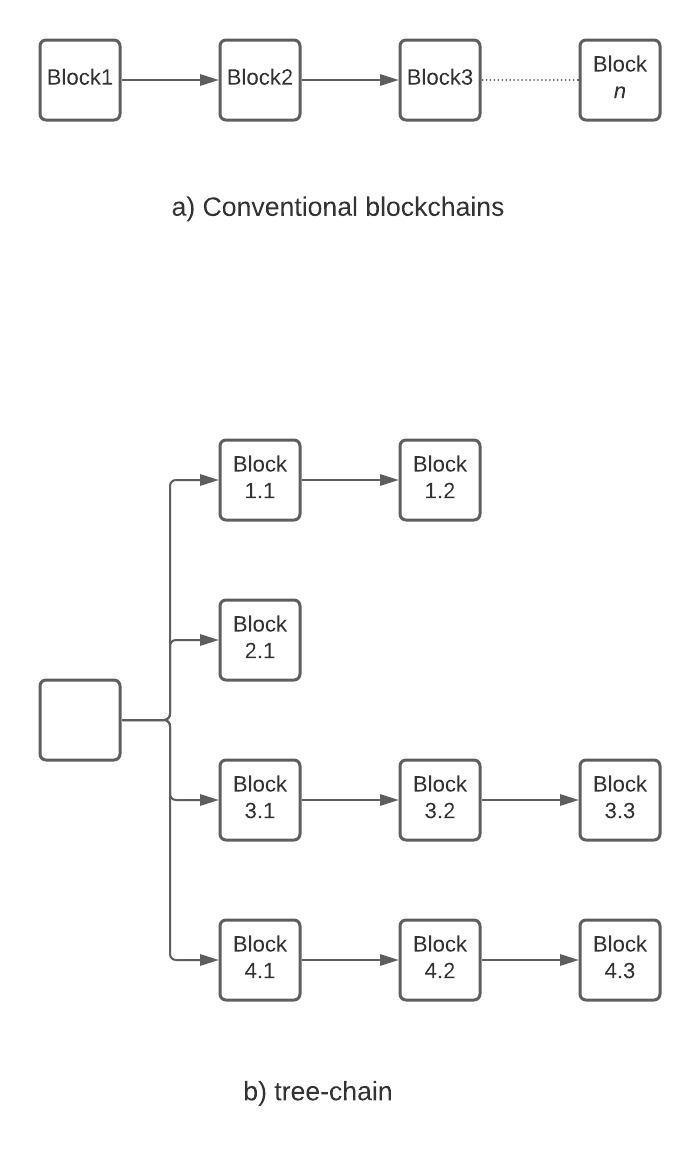}   
		\caption{A high level view of a) conventional blockchains and b) Tree-chain.}
		\label{fig:conventionalchains-vs-tree-chain}
	\end{center}
\end{figure}

As shown in Figure \ref{fig:conventionalchains-vs-tree-chain}, Tree-chain embraces the concept of forking where each ledger is managed by a particular validator, i.e., each ledger contains transactions that all fall within the same consensus code. Each ledger is maintained by a particular validator. Depending on the weight of their PK, the validators are allocated a \textit{Consensus Code Range}, which is the most significant characters of the hash function output. Each validator only commits transactions whose hash falls within its corresponding consensus code range.  Unlike tree-chain, Vericom is not a validator selection algorithm. Instead, Vericom focuses on routing and verifying the already generated blocks and transactions.\par

Another basic building block   we partially employed in Vericom is  the routing algorithm proposed in  our earlier work  \cite{dorri2019spb}.   We introduced an anonymous routing method that routes  transactions from the source node to the destination on the basis of PK. Designated nodes in the network form a backbone network. Each backbone node receives transactions where the hash of the transaction destination starts with specific characters. The destination nodes  join the backbone node that is responsible for managing their transactions by sending a join request transaction to the backbone node.  The backbone nodes route transactions  based on the PK of the destination.\par  

In this paper, we employ the concept of backbone network, however, the backbone network is responsible for managing all blockchain traffic, including blocks and transactions, and introduces a new method to define the destination of the incoming traffic based the hash function output of the traffic.  The backbone nodes do not broadcast the traffic, instead they multicast the traffic flow  to a verifier set, which in turn reduces the packet overhead and bandwidth consumption.  \par 

Having discussed the background information, we next outline the details of Vericom.

\subsection{Vericom} \label{sec:vericom-core}

\begin{figure}
	\begin{center}
		\includegraphics[width=11cm ,height=10cm ,keepaspectratio]{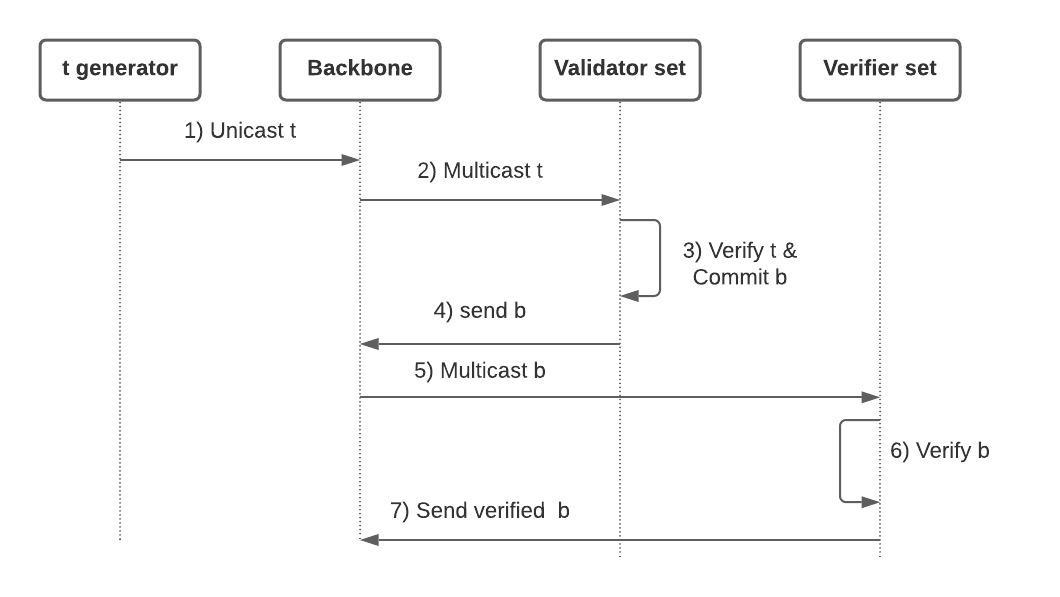}   
		\caption{A high level view of information flow in Vericom (\textit{t} refers to a transaction and \textit{b} refers to a block.) }
		\label{fig:vericom-trans-flow}
	\end{center}
\end{figure}

\begin{figure}
	\begin{center}
		\includegraphics[width=8cm ,height=8cm ,keepaspectratio]{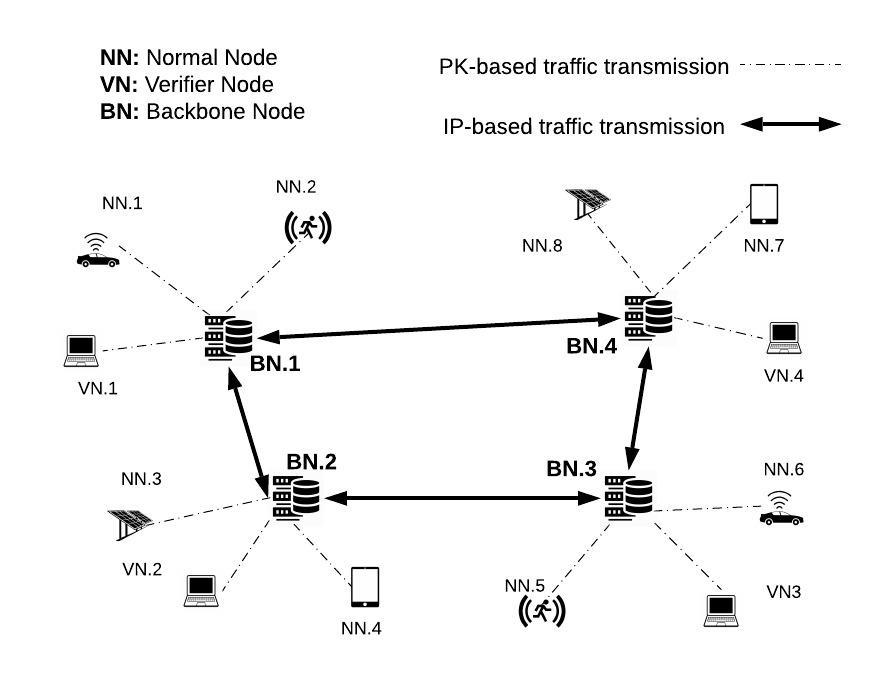}   
		\caption{A high level view of  Vericom.}
		\label{fig:overal-design}
	\end{center}
\end{figure}

Vericom  introduces a distributed yet scalable blockchain  by  shifting from broadcasting  to dynamic and randomized multicasting of the traffic flow.  A high-level picture of Vericom is shown in Figure \ref{fig:overal-design} and a summary of the core steps is given in Algorithm \ref{algo:vericom}. As Vericom logic is distributed across several network entities, the first entity mentioned in each line in Algorithm \ref{algo:vericom} is the entity that conducts the action explained in the line. The verifiers of   a particular transaction/block are identified based on the output of the hash of the traffic content which ensures randomized and dynamic  verifier selection.  This in turn reduces the communications overhead and computational overhead for verifying new blocks and transactions. Verifying a  transaction/block typically involves matching the PK with the corresponding signature.  Vericom consists of two layers namely: i) transmission layer: where the traffic is multicasted to a dynamically selected group of recipients, and ii) verification layer: where a set of nodes are randomly selected to verify  transactions or blocks.  Vericom differentiates between  verifiers and validators. A verifier is a node   that verifies  a newly generated block, while validator  refers to a  node that first verifies newly generated  transactions then commits them in the blockchian by following Tree-chain consensus algorithm. Figure \ref{fig:vericom-trans-flow} depicts a high-level view of Vericom traffic flow.    Algorithm \ref{algo:vericom} identifies the flow in which the traffic passes these layers which is further explained in the rest of this section. 

\par 

\begin{algorithm}[tb!]
	\caption{A summary of Vericom (N: normal node, BN: backbone node, VN: verifier node).}\label{algo:vericom}
	\begin{algorithmic}[1]
		\State \textbf{N}: Send \textit{t\textsubscript{n}} to BN \Comment{Transmission layer}
		\State \textbf{BN:} Identify the verifiers based on hash 
		\State \textbf{BN:} Multicast  \textit{t\textsubscript{n}} to VN 
		\If{ VN matches  \textit{t\textsubscript{n}}.PK with \textit{t\textsubscript{n}}.Sign }\Comment{Verification layer}\\
		\textit{t\textsubscript{n}} is verified
		\EndIf
		\State \textbf{VN:} Send 	\textit{t\textsubscript{n}} to BN \Comment{Transmission layer}
		\State \textbf{BN:} Send 	\textit{t\textsubscript{n}} to the Validator (Val)
		\State \textbf{Val:} Commit    \textit{t\textsubscript{n}} to a new block (B)  
		\State \textbf{Val:} Send B to BN \Comment{Transmission layer}
		\State \textbf{BN:} Multicast B to the verifiers 
		\State \textbf{VN:} Verify B \Comment{Verification layer}
		\State \textbf{VN:} Send verified B to BN \Comment{Transmission layer}
		\State \textbf{BN:} Broadcast B
		
	\end{algorithmic}
\end{algorithm}

As an example scenario, consider the network shown in Figure \ref{fig:overal-design}. Backbone Node.1 (BN.1), BN.2, BN.3, and BN.4 form the backbone network. Assume the  consensus code allocated to each of the Validator Nodes (VNs) is as in Table \ref{tab:consensus-code}. Normal Node.1 (NN.1) generates a transaction whose hash is "K23HQ" and sends to its corresponding backbone node, i.e. BN.1 (line 1, Algorithm \ref{algo:vericom}). The transaction is multicasted to: i) the validator set (line 2) which in this example are the main validator (selected based on the consensus code allocations and the hash of the transaction, VN.3 in this scenario, and ii) the validator that is allocated to the next consensus code range of the main validator, i.e., VN.4 in this scenario.  BN.1 routes the transaction to  BN.3 and BN.4 using their corresponding IP addresses (line 3). Each BN populates  and updates a routing table based on conventional IP-based routing algorithms that is used to route traffic in backbone network.  Upon receipt of the transaction, BN.3 and BN.4 send the  transaction to VN.3 and VN.4 to be verified (lines 4\&5) and committed in the blockchain. Depending on the application, the verification of a transaction may involve different steps. Recall that we assume Tree-chain is employed as the underlying consensus algorithm, thus the main  validator  commits  the transaction in the blockchain (line 8, the validator skips lines 6\&7 as the same nodes that verify  transactions commit them in blockchain). After generating a new block, the main validator  sends the block to the backbone node (line 9) which is then multicasted to the verifier set (line 10). After verifying the new block, the verifiers send it to the backbone nodes to be broadcast in the network (lines11-13).   In conventional blockchains for a transaction to be verified and committed two broadcasts shall happen (broadcasting the transaction and the block). Vericom relies on multicasting for transaction and block verification, however, eventually the block needs to be broadcast to be stored by the nodes. The storage process is beyond the scope of Vericom and thus we leave that for  future work. We discuss the details of the outlined process in the rest of this section.\par

\begin{table}\caption{Consensus code allocation for scenario  in Figure \ref{fig:overal-design}.}\label{tab:consensus-code}
	\begin{center}
		\begin{tabular}{ |c|c| } 
			
			\hline
			\textit{\textbf{Validator ID}} & \textit{\textbf{Consensus code}} \\\hline
			VN.1 & [0-9]  \\ \hline
			VN.2 & [A-H] \\ \hline
			VN.3 & [I-P]  \\ \hline
			VN.4 & [Q-Z]  \\ \hline
			
		\end{tabular} 
	\end{center}	
\end{table}

\subsubsection{Transmission layer}\par 
Unlike conventional blockchains where traffic is broadcast, in Vericom, the traffic is routed by the backbone nodes and then multicasted to intended recipient groups that are dynamically identified based on the verification layer (discussed later in this section). To achieve this goal, we introduce a backbone network along with a PK-based multicasting. Highly stable and resourceful nodes  in the network jointly form a \textit{backbone} network that is the core for traffic management. The backbone network receives  traffic from all  IoT nodes, e.g. solar panels or IoT devices in a smart home, and delivers it to the intended destination.  The formation of the backbone  depends on the level of trust to the backbone nodes. We consider two scenarios trusted backbone nodes and untrusted backbone nodes, which we discuss separately below. \par 

\textbf{Trusted Backbone Nodes:} In this scenario, the backbone nodes are trusted, e.g., the servers provided  by the Internet Service Providers (ISPs) or the government. This is similar to the Internet backbone network  that manages the main Internet traffic flow. In conventional blockchains the traffic is technically broadcast in an overlay network, while in the lower layers, the traffic is   routed by the \textit{trusted} Internet backbone nodes (as with any other  Internet traffic).  Vericom aims to remove the overlay network in conventional blockchains and  incorporate   the backbone network concept  in the  blockchain design	 to   reduce  delay and packet overhead. As outlined earlier, even the existing pure distributed blockchains rely on trusted Internet backbone nodes to relay traffic, thus, the trust level to the backbone nodes does not impact the distributed nature of the blockchain. The reason is that while communication traverses the backbone, the routing decisions are made independently of the backbone. Conventionally, transactions and blocks are broadcast to all participants. Vericom replaces broadcast with dynamic multicast to a randomly selected set of nodes based on hash outputs, which maintains independence of routing decisions from the backbone and avoids any centralization of trust. 	Apart from the underlying  transmission layer,  blockchain tasks are still conducted in a distributed manner as outlined later in verification layer. \par 

The trusted backbone nodes  communicate to form the backbone network. We assume an authorized organization, e.g., the government or the Internet backbone network provider, connects  the potential backbone nodes. Vericom routes traffic based on the hash function output of the traffic as outlined in the verification layer.

\textbf{Untrusted Backbone Nodes:} In this scenario, we assume that the backbone nodes are not trusted, neither to other backbone nodes nor to the network participants. This in turn makes security challenging as a malicious backbone node may attempt to drop incoming/outgoing traffic. Note that as blockchain transactions are sealed using asymmetric  encryption, modification of the transaction/block content by the malicious backbone nodes  is not possible. To enhance the security of Vericom in the presence of untrusted backbone nodes, we employ neighbor monitoring algorithms (such as in \cite{hai2008detecting}) where the behavior of each backbone node is monitored by its neighbors which in turn can detect malicious activities. We introduce a random and unpredictable  method to select neighbors for monitoring the behavior of each node to enhance the security as compared with conventional neighbor monitoring methods. The neighbors will also be incentivized to participate in backbone network by receiving a fee (as discussed later in this section).

The backbone nodes receive traffic from the source IoT nodes  and route them to the proper destination, that is identified based on the verification layer (see Algorithm \ref{algo:vericom}). To receive traffic, the nodes  shall join  at least one backbone node, otherwise,  they  will fail to receive traffic and thus will be isolated from the network.  \par

Each  validator or verifier   orders the list of backbone nodes based on the delay experienced to reach them and sends a join request to the  backbone node with minimum delay. The backbone nodes have limited resources and thus can serve limited number of nodes (depending on the amount of available resources and the volume of traffic in the network). To avoid queuing and thus reduce communication  delay, each backbone node  accepts  join requests only from a  particular number of  nodes. Once the maximum number of nodes join, the backbone node rejects the join requests from the new nodes. The nodes then send the join request to the next backbone node in their list. This ensures the minimum delay in routing traffic in the network. Once all nodes  join the backbone network, each backbone node broadcasts the PK of its corresponding nodes to the rest of the backbone nodes. The  backbone nodes maintain a routing table that stores the list of nodes    associated with other backbone nodes and their corresponding  role (i.e., validator or verifier). During a particular time-interval, known as route update interval (RUI),  the backbone nodes broadcast an update packet that contains the updated list of  nodes connected to the backbone node.  This in turn ensures that the backbone nodes can update the routing information in case of any change to the underlying nodes. To reduce  packet overhead, each backbone node only generates an RUI if changes happened to the underlying nodes list since the last RUI (or initialization). As an example the routing table of BN.1 in Figure \ref{fig:overal-design} is as shown in Table \ref{tab:routing-table}. To reduce the size of the routing table, each BN may only store the next hop in the path to reach a particular verifier (which is similar to routing tables in the Internet).  The backbone nodes employ conventional routing algorithms such as OSFP [RFC 2328] to decide on the next hop node toward the destination. \par

\begin{table}\caption{The routing table of BN.1 in Figure \ref{fig:overal-design}.}\label{tab:routing-table}
	\begin{center}
		\begin{tabular}{ |c|c| } 
			
			\hline
			\textit{\textbf{validator ID}} & \textit{\textbf{Next hop}}  \\\hline
			VN.4 & 4 \\ \hline
			VN.2 & 2  \\ \hline
			VN.3 & 2  \\ \hline

		\end{tabular} 
	\end{center}	
\end{table}

The backbone  nodes  dedicate communication and computational  resources to  manage the traffic flow that in turn incurs monetary costs. To incentivize nodes to join the  backbone network, we introduce  a \textit{Traffic Management Fee (TMF)}. TMF  is paid by the  validators based on the total number of blocks they generated during an epoch time known as $ \Theta$. In our setting, $ \Theta = consensus\_period$ (see Section \ref{sec:background}).  At the end of a $ \Theta$ each validator calculates TMF as:\par 
\textit{TMF =  TF * ledger\_length }\par 
where \textit{Traffic fee (TF) }  is the cost for forwarding the traffic related to a single block and is defined by the blockchain designers. The size of the block and the bandwidth cost are two key factors that impact TF value.    \textit{Ledger\_length }is the  total number of the blocks each validator has generated  during $ \Theta$.  Recall that we base Vericom on tree-chain where  each validator stores blocks in its own ledger. The validator then pays  TMF to a \textit{Traffic Accounting (TA)}  smart contract. TA is hard coded by the blockchain designers in the first block in the blockchain that is known as the \textit{genesis block}. \par

TA collects TMF from the validators and then equally distributes between the backbone nodes. Upon receipt of the payment from the validator, TA verifies if the validator has paid the right amount by calculating the TMF  and matching with the received fund. If a validator pays less or no fee to the TA contract, the TA will notify the rest of the network by broadcasting a transaction. The validator will be denied from joining the validators in the next consensus period round until payment is made. \par

As evident from the above discussion, the only penalty for a validator that pays no/less fee is to prevent them from functioning as validator in future. Note that TMF is small and the benefits of being a validator are much higher than the TMF (as the validators collect transaction fees).  An alternative option is to block specific amount of money from each validator initially and deduct from the blocked money in case the validator fails to pay the fee. \par

As outlined above, the transmission layer receives the traffic from the source nodes and multicast to dedicated nodes that are identified by the \textit{verification layer}.

\subsubsection{Verification layer}\par 
The verification layer  aims to identify the destination of the traffic received by the backbone nodes. This reduces the computational overhead associated with transactions/block verification, compared to conventional blockchains, where all nodes must verify the new transactions or blocks. The verification of new transactions or blocks typically involves matching the signature with the PK of each transaction.  To protect against malicious nodes that may falsely claim a fake transaction/block as valid, the destination nodes are selected randomly, dynamically, and uniquely for each transaction/block. The destination is  either  i) validator set that verify and commit new transactions in the blockchain by following the consensus algorithm, or ii) verifier set that is a group of nodes that verify newly generated blocks.   Both these sets are randomly and dynamically selected from the same pool of nodes in the network based on the hash of the traffic as outlined below.    \par

To identify the validator/verifier set corresponding to each transaction/block, Vericom relies on the hash function output  of the traffic. Each potential character in the hash function output is allocated  a particular weight that is identified in a \textit{Weight Dictionary (WD)} an example of which is shown in Table \ref{tab:KWMExample}.  During the bootstrapping, the participating nodes in the blockchain (referred to as \textit{PN}) that are interested to function as validator/verifier  broadcast a \textit{validator interest transaction} that contains their PK. The nodes shall broadcast the interest transaction within a particular time frame referred to as $\gamma$, defined by the network designers.   To ensure consistency among the participating nodes, during the network setup, a \textit{Validation Range Distributor (VRD) }  smart contract is deployed in the genesis block, i.e., the first block in the blockchain. \textit{Validation  Range (VR)}  is a range of  \textit{Most Significant Character (MSCh)} of the hash function output.  To communicate with VRD, the participating nodes shall populate the address of the VRD, i.e., the hash of its content, as the destination in their transaction.  The   interested nodes  send their PK to the VRD. \par

\begin{table}[h]
	\caption{An example of WD.}\centering
	\begin{tabular}  { | p {2.5 cm} |   p {2 cm}|}
		\hline
		\textbf{ $ \alpha $}     & \textbf{$ \omega(\alpha)  $} \\\hline
		a-z & 0-25  \\\hline
		A-Z & 26-51  \\\hline
		0-9  & 	52-61 \\\hline
	\end{tabular}
	\label{tab:KWMExample}
\end{table}

At the end of   $\gamma$, the VRD smart contract  starts calculating a Key Weight Metric (KWM) corresponding to the each received PK.  KWM is calculated as follows (\textit{k} is the size of the hash funciton output):\\

$  \displaystyle\sum_{i=1}^{k} fw(\alpha\textsubscript{i})    $

where $fw(\alpha)$ is the final weight of  $\alpha$ which is calculated as: 

\[ fw(\alpha)=
\begin{cases}

w(\alpha)    & \text{if } r = 0   \\
w(\alpha) *(0.2^r) &   \text{if }  r > 0
\end{cases}
\]
where \textit{r} is the number of times that $\alpha$ is repeated in the hash function output. This in turn ensures that the final KWM corresponding to each PK is unique.  The VRD then creates a descending list of the PKs based on the KWM values and  allocate  a particular VR. 	 The size of  VR is defined as follows: \\
$ \lceil{\frac{number\_of\_validators}{62}} \rceil$ \\
where 62 is the total number of possible values in each byte of the hash output which is from the following range:  \{0,..,9,a,...,z,A,...,Z\}. The above division may have fractions. In that case, the first node in the list will be allocated a larger VR to accommodate all values. As an example, in a network with 10 nodes, the size of the VR allocated to the first  node is 8 and the size of VR for other nodes is 6.   The VR is allocated to the validators based on the position of PK of the validator in the KWM list, e.g., the validator with the highest value of KWM is allocated to the first VR. The validators form a Distributed Hash Table (DHT) that includes the PK of each validator and its corresponding VR that is used to identify the corresponding VR to each validator during for routing traffic.  \par

When a backbone node receives traffic from the source node, it decides on the verifier/validator set by  evaluating the output of the  hash of the traffic, i.e., block or transaction content. Each  set contains a main node, which acts as the leader of the nodes in the  set, and the sub nodes which are the nodes that monitor the behavior of the main node. The backbone node first must identify the main node in the  set. For transaction \textit{t\textsubscript{i}},  the main validator  is the validator whose VR covers the MSCh of  \textit{h(t\textsubscript{i})} where \textit{h(x)} represents the hash function output of \textit{x} (Step 2 Figure \ref{fig:vericom-trans-flow}).   Once the main validator is identified, the backbone nodes add \textit{n} successors and predecessors   of the main validator  to the validator set as sub validators defined in the DHT corresponding to the VR allocation.  $  1 <= n <= N/4 $ where \textit{N} refers to the total number of validators in the network. The upper bound value for \textit{n} is identified in a way to ensure there will be no overlap between validator set and verifier set (see discussion in the rest of this section).    Successors and predecessors  are defined as the nodes that are immediately after and before the main validator in DHT according to the KWM.    The value of \textit{n} depends on the application. Larger \textit{n} increases security to colluding malicious nodes in the validator sets, but also increases computional overhead for verifying and committing  new transactions.  The validator set first has to verify the transaction that involves  matching the  PK of the transaction  with the corresponding signature (step 3, Figure \ref{fig:vericom-trans-flow}). Depending on the application, other steps  might also be involved, e.g. if a transaction is chained to a previous transaction, the verifier shall verify the existence of the previous transaction. After verification, the main validator   commits   the transaction  in the blockchain into a new block (following Tree-chain algorithm).  The sub validators   monitor the behavior of the main validator and inform the rest of the network in case any malicious activity, e.g., not storing new transactions or storing fake transactions, is detected. In case a different consensus algorithm other than Tree-chain is employed, the validators will add the transactions that are already verified by the verifier set to the pool of pending transactions.  \par

\begin{figure}
	\begin{center}
		\includegraphics[width=8cm ,height=8cm ,keepaspectratio]{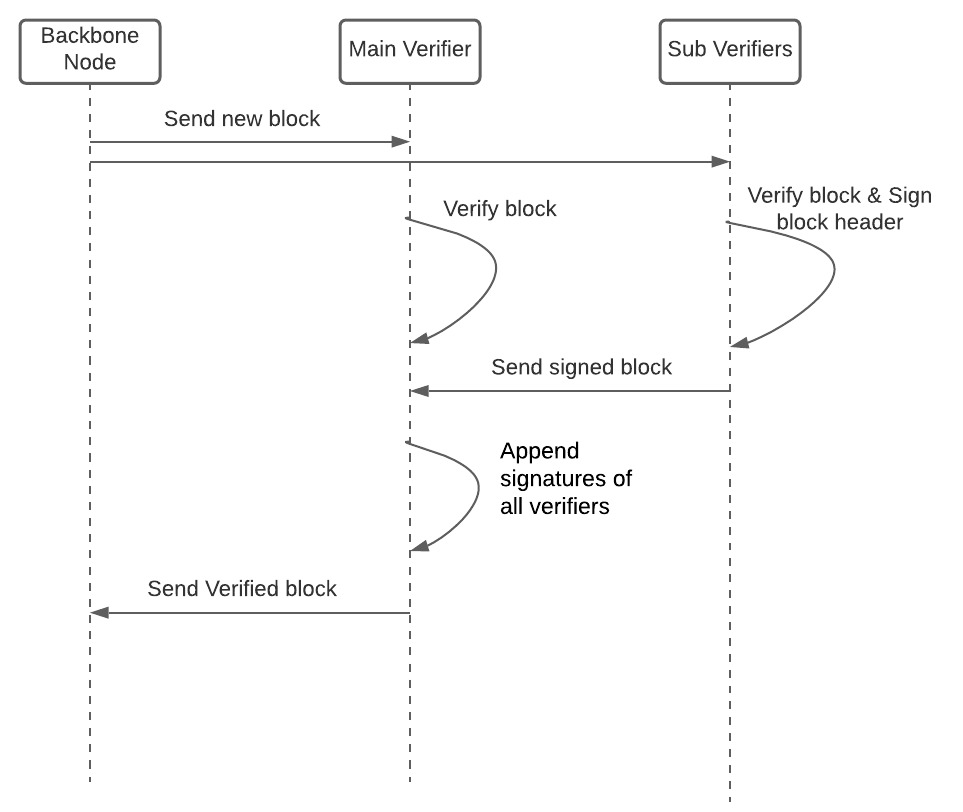}   
		\caption{The process of verifying new blocks.}
		\label{fig:block-verification}
	\end{center}
\end{figure}

Once the new block is committed, the validator sends the block to its corresponding backbone node to be verified (step 4, Figure \ref{fig:vericom-trans-flow}). The block must be signed by the nodes in the verifier set to be considered as a valid block in the blockchain. Figure \ref{fig:block-verification} shows the steps involved  in verifying new blocks. Similar to validator set, the verifier set  is identified based on the hash of the block. The verifier set consists of the main verifier and subverifiers which are \textit{m} successors and predecessors   of the main verifier.  To enhance the security of Vericom against malicious nodes that may attempt to commit fake blocks and verify them as true blocks, the underlying  nodes in the validator  set shall always be different from the verifier set. In case the validator and verifier set have overlapping nodes, the main verifier will be: \par

\[ Main Verifier =
\begin{cases}
2n      & \quad \text{if } n > m  \\
2n+m  & \quad \text{if } n<= m  
\end{cases}
\]

This ensures that there will be no overlapping nodes in two sets. Once the nodes in the verifier set received the new block ((step 5, Figure \ref{fig:vericom-trans-flow})), they first verify the block and then sign the block and sends to the main verifier. The verification of the block involves verifying the underlying transactions and the block header (step 6, Figure \ref{fig:vericom-trans-flow}). The transaction verification is as outlined earlier. The block header verification involves matching the PK of the block generator with the corresponding signature. Next, the verifier checks if the hash of the block content falls within VR of the block generator. The main verifier adds the signatures of all the nodes in the verifier set to the block and multicast to the other validators  in the network  to be stored in the blockchain (step 7, Figure \ref{fig:vericom-trans-flow}).  The validators may accept the signed block  by the verifier set without verifying the underlying transactions.   All nodes in the blockchain still can verify the traffic confirmed by the verifier set which in turn will detect any malicious activity (see Section \ref{sec:evaluation}).   \par

In summary, the verification layer ensures that each transaction/block is verified twice by randomly selected nodes:    first by the transaction validator set and second by the block verifier set.

\section{Evaluation and Discussion}\label{sec:evaluation}
In this section we provide qualitative security analysis as well as quantitative performance evaluation. 

\subsection{Security and Privacy  Analysis}

\textbf{Threat model:} The malicious node can be any node in the network. In case of trusted backbone nodes, it is assumed that the backbone nodes function honestly and employ security safeguards that protects them against being compromised. The malicious node may drop the received traffic, generate fake transactions or blocks, and falsely claim that a transaction is valid. The malicious node may collaborate with other nodes to increase the chance of a successful attack.

\textbf{Privacy:}  Vericom does not impact the anonymity level of the transactions stored in the blockchain. However, in conventional blockchains the transactions are broadcast thus the underlying nodes may not record the IP address of the  underlying IoT nodes. In Vericom, the participating nodes shall record the IP of the backbone nodes which may potentially reveal information about the actual location of the backbone nodes. Note that the backbone nodes  only forward  traffic. If a backbone node wishes to generate a transaction, it employs a PK similar to other nodes which protects the anonymity of the node. \par 

Recall that the backbone nodes multicast traffic to the intended group of recipients. This may potentially require the backbone node to record the IP address of the nodes in the verifier set. To prevent the backbone nodes from tracking the location based on IP, the verifiers may hide their IP using TOR browser \cite{Tor}.\par

\textbf{Security:} We discuss three security attacks:\par 

\textit{False verification attack:} In this attack, an IoT node or a group of nodes in the verifier set collaborate to mark a fake transaction or block as valid (step 3 \& 6, Figure \ref{fig:vericom-trans-flow}). Recall from Section \ref{sec:architecture} that Vericom limits the number of nodes that verify a transaction or block to a verifier set. If a single verifier attempts to make false claim,  other verifiers will detect the malicious activity. The verifiers then broadcast a transaction to inform the rest of the network of the malicious behavior. The fake transaction is also broadcast to the network that enables the rest of the IoT nodes to verify the claim of the verifier set.\par

It may be possible that all the nodes in the verifier set collaboratively attempt to store fake transactions in the blockchain. As outlined in Section \ref{sec:architecture}, once the main verifier commits the transaction in the blockchain, the block is sent to a new set of randomly selected nodes for verification which will detect the fake transaction (lines 9-11, Algorithm \ref{algo:vericom}). If the new selected verifier set  are malicious, they accept the block and mark it as valid.  Vericom is designed for IoT that comprises  millions of nodes, thus, it is expected that a large number of nodes will participate in the blockchain management which in turn makes it challenging for the malicious nodes to compromise all nodes in a randomly selected verifier set.  Recall that any node may attempt to verify the transactions and blocks that are already verified by the  verifier set. Such nodes will detect the malicious behavior. Given the large scale of  IoT, it is highly likely that at least one node will allocate resources to verify transactions, this can be the SPs to ensure secure and safe services. 

\textit{Fake transaction storage attack: } In this attack, the malicious node attempts to store a fake transaction in the blockchain. The verifier of each transaction is identified based on the hash of the transaction content which is random and unpredictable. The malicious node might have control over (or collaborate with)  another verifiers. In such case, if the transaction hash falls within the consensus code range managed by the malicious nodes, they can mark the fake transaction as valid. The validators  attempt  to store the verified transactions in the blockchain. \par

New blocks are verified by  a randomly selected  verifier set which will detect  the fake transaction. In the worst case, if the nodes in the verifier set  collaborate with the malicious node, the fake transaction   will be broadcast as a valid transaction. Note that other nodes in the network can still verify the transctions and may detect the fake transaction. If so, the transaction ID is broadcast to the network the the malicious nodes, including the verifiers, will be removed from the verification layer. It also worth noting that Vericom is designed for IoT where the number of verifiers is expected to be large. Plus, the verifiers are always selected randomly in an unpredictable manner, thus, it is hard for the malicious nodes to control the block generator and the verifier set. \par

\textit{Dropping attack:} In this attack, a backbone node drops the incoming traffic to prevent blockchain participants from receiving service (applicable only in untrusted backbone node's scenario). Recall from Section \ref{sec:architecture} that in the untrusted backbone node scenario, a group of nodes collaboratively manage the traffic where the main node forwards the traffic and other nodes monitor the behavior of the main node which in turn enable them to detect any malicious activity.  In case of a collaborative attack, the IoT nodes will not receive any service, thus will inform the rest of the network. The backbone nodes then will reconstruct the backbone network  and prevent  the malicious nodes from joining the network.

\subsection{Performance Evaluation}
In this section, we study the performance of Vericom. We implemented Vericom using NS3 \cite{NS3} incorporated with crypto++ security  library to implement security features. During the evaluation, a total of 1000 transactions have been generated by the participating nodes in the blockchain. We set the size of the verifier set as 3 in our experiments. Note that  Vericom employs tree-chain \cite{dorri2020tree} as the underlying consensus algorithm. Both tree-chain and Vericom incorporate fundamental changes to the conventional blockchains, thus we were unable to use the existing blockchain simulation platforms, such as Hyperledger \cite{Ethereum} or Ethereum \cite{HyperledgerFabric}. We compare Vericom performance with a \textit{baseline} scenario which is similar to conventional blockchains where all transactions and blocks are broadcast and verified by all participating IoT nodes. As our focus is on maintaining distribution for IoT security, we refrain from comparing against hierarchical, clustered, or sharded blockchains, as they intrinsically recentralise security and trust.  We studied the following metrics:

\begin{itemize}
	\item Packet overhead:  This is the cumulative packet overhead incurred on the participating IoT nodes   and is measured by summing  the size of the received traffic  by these nodes. 
	
	\item Verification processing time: This is the processing time taken from a verifier  to verify the transactions and blocks in the blockchain. We disregard the processing time associated with other blockchain-related tasks, e.g., consensus,  as Vericom is not impacted by those.

	\item Delay: This is the time taken for sending a transaction from IoT node \textit{A} to IoT node \textit{B}.  
	
	\item Vericom overheads: This is the extra overheads incurred by Vericom and includes: i) the increased block size  to include the verifier set PK and signature, and ii) the size of the routing table.  
	
\end{itemize}

We first evaluate   the scaling performance metrics  as a function of the key network parameters. Table  \ref{tab:scaling-performance} outlines the  results. In Vericom, the packet overhead and delay in reaching another node depends on the number of hops in the backbone network while in baseline such overheads depend on the number of IoT nodes in the network which is significantly greater than Vericom. Similarly, the verification processing time in Vericom relies   only on the number of nodes in the verifier set while the baseline is impacted by the number of participating IoT nodes in the network.  In Vericom the packet overhead scales linearly with the number of hopes compared to  quadratic overhead for conventional blockchain networks. In the worst case scenario where the nodes are connected linearly, i.e., in a chain structure, the packet overhead in Vericom will be \textit{O(N)}  as the packets should travel through all nodes.

\par 

\begin{table}\caption{Evaluating the scaling performance.}\label{tab:scaling-performance}
	\begin{center}
		\begin{tabular}{ |c|c|c| } 
			
			\hline
			\textit{\textbf{Metric}} & \textit{\textbf{Vericom}} & \textit{\textbf{Baseline}}\\\hline
			Packet overhead & O(H) & O(N$^2$)  \\ \hline
			Verification processing time & O(V) & O(N) \\ \hline
			Delay & O(H) & O(N)  \\ \hline
		\end{tabular} 
	\end{center}	
	H: Number of hops in the backbone network.\\
	N: Number of IoT nodes in the blockchain.\\
	V: Size of the verifier set. \\
	
\end{table}

Having discussed the scaling performance evaluation, we next evaluate  the performance of Vericom based on the simulation output.\par 

\begin{figure}
	\begin{center}
		\includegraphics[width=8cm ,height=8cm ,keepaspectratio]{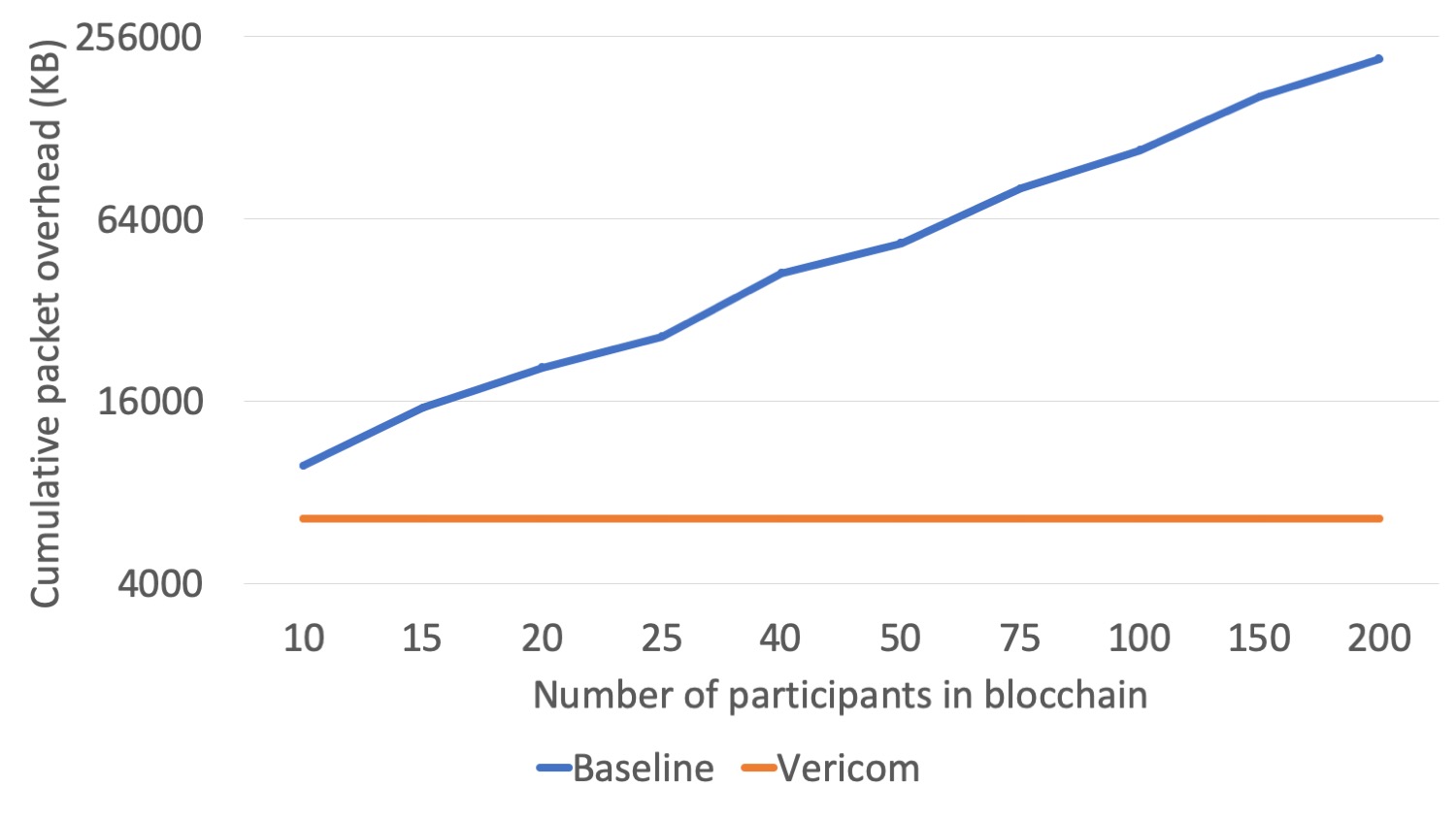}   
		\caption{The cumulative packet overhead.}
		\label{fig:ev:packet-overhead}
	\end{center}
\end{figure}

\textit{Packet overhead:}  The simulation results to evaluate the packet overhead are outlined in Figure \ref{fig:ev:packet-overhead}. We increase the number of underlying IoT nodes from 10 to 200 while the backbone network remains static as 20 nodes.  As evident from the results, Vericom packet overhead remains constant  as the number of IoT nodes  increases at around 6,000 KB, while the baseline packet overhead significantly increases reaching from 10,000 KB to around 216,000 KB. The main reason is that in Vericom the packets travel only among the backbone network and the verifier set and thus the increased in the number of underlying IoT nodes does not impact the general packet overhead, resulting in a reduction of packet overhead of up to nearly two orders of magnitude in this scenario over the baseline approach.  However, in the baseline, the packets must be broadcast to the whole network and thus, as shown in Table \ref{tab:scaling-performance}, the packet overhead increases  as  the number of  participating IoT nodes increases.  \par

\begin{figure}
	\begin{center}
		\includegraphics[width=8cm ,height=8cm ,keepaspectratio]{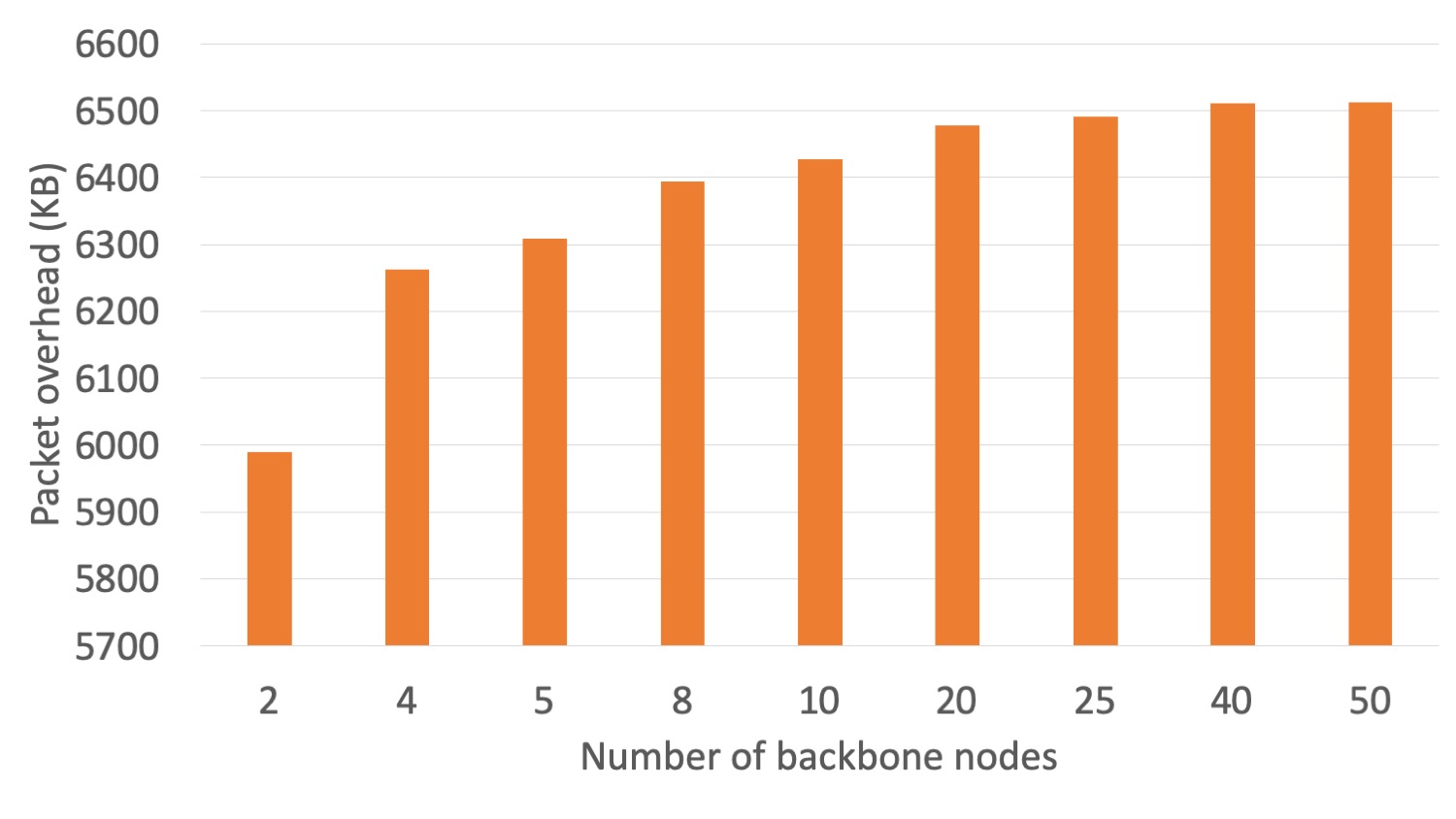}   
		\caption{Studying the impact of backbone nodes on packet overhead. }
		\label{fig:ev:packet-overhead-backbone}
	\end{center}
\end{figure}

Vericom packet overhead is largely impacted by the number of backbone nodes. Next, we change the number of backbone nodes in the network while the number of the underlying IoT nodes remains constant as 200 nodes.  The new backbone nodes  connect to a randomly selected existing backbone node.  The source and the destination of the blockchain traffic are  selected randomly that  ensures  the structure of the backbone network changes dynamically. The simulation results are shown in Figure \ref{fig:ev:packet-overhead-backbone}. The packet overhead increases from around 5,900 KB  with 2 backbone node to 6,500 KB  with 50 backbone nodes. Recall from Table \ref{tab:scaling-performance} that Vericom packet overhead depends on the number of hops that the packets travel to reach a destination, thus the increased packet overhead for larger number of backbone nodes is relatively small.

\begin{figure}
	\begin{center}
		\includegraphics[width=8cm ,height=8cm ,keepaspectratio]{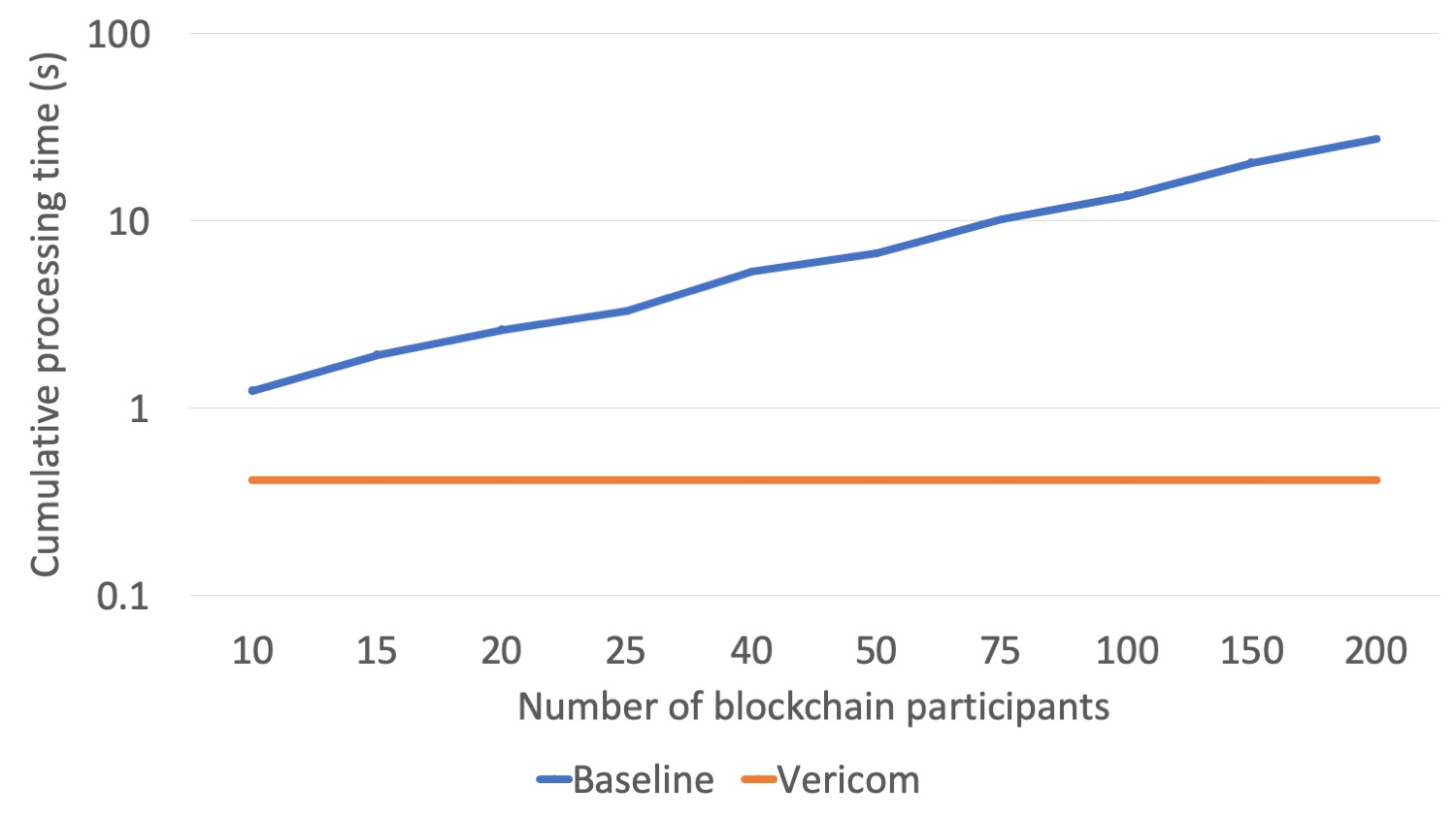}   
		\caption{Studying the cumulative processing time. }
		\label{fig:ev:processing-overhead}
	\end{center}
\end{figure}

\textit{Verification processing time:}  Figure \ref{fig:ev:processing-overhead} compares the processing time in the baseline and Vericom to verify new transactions/blocks.  Despite the fact that the underlying verifiers dynamically change, the number of nodes that verify new transactions/blocks, i.e., the size of the verifier set, always remains constant, that is 3 in our experiment, which results in a constant processing time in Vericom as the number of IoT nodes increases.  However, increasing the number of IoT nodes in the baseline, increases the processing time as all the nodes  verify new transactions/blocks.

\begin{figure}
	\begin{center}
		\includegraphics[width=8cm ,height=8cm ,keepaspectratio]{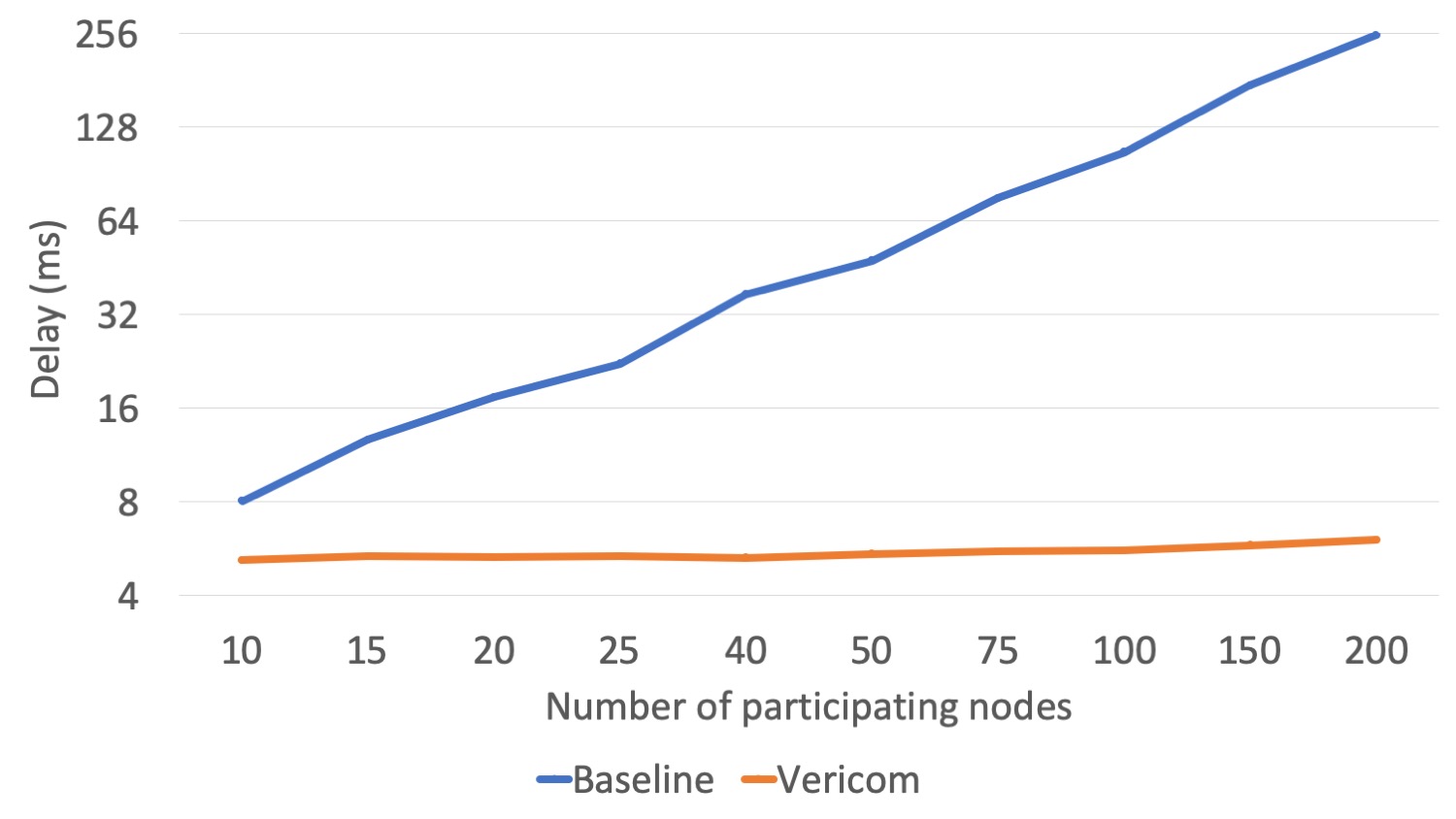}   
		\caption{Studying the communication delay. }
		\label{fig:ev:delay}
	\end{center}
\end{figure}

\textit{Delay:} The simulation results to evaluate the delay in communications between two nodes are represented in Figure \ref{fig:ev:delay}. We randomly select two  nodes in the network to measure the delay while increasing the number of blockchain participants from 10-200 nodes. The delay in reaching another node in Vericom  is in the range of  [5-6] milliseconds while delay in the baseline is in the range of   [8-253] milliseconds. Recall from Section \ref{sec:architecture} that Vericom routes packets among the backbone nodes to reduce the delay while in baseline, the packets are broadcast till reaching the destination which in turn significantly increases the delay in communications. \par

\begin{figure}
	\begin{center}
		\includegraphics[width=8cm ,height=8cm ,keepaspectratio]{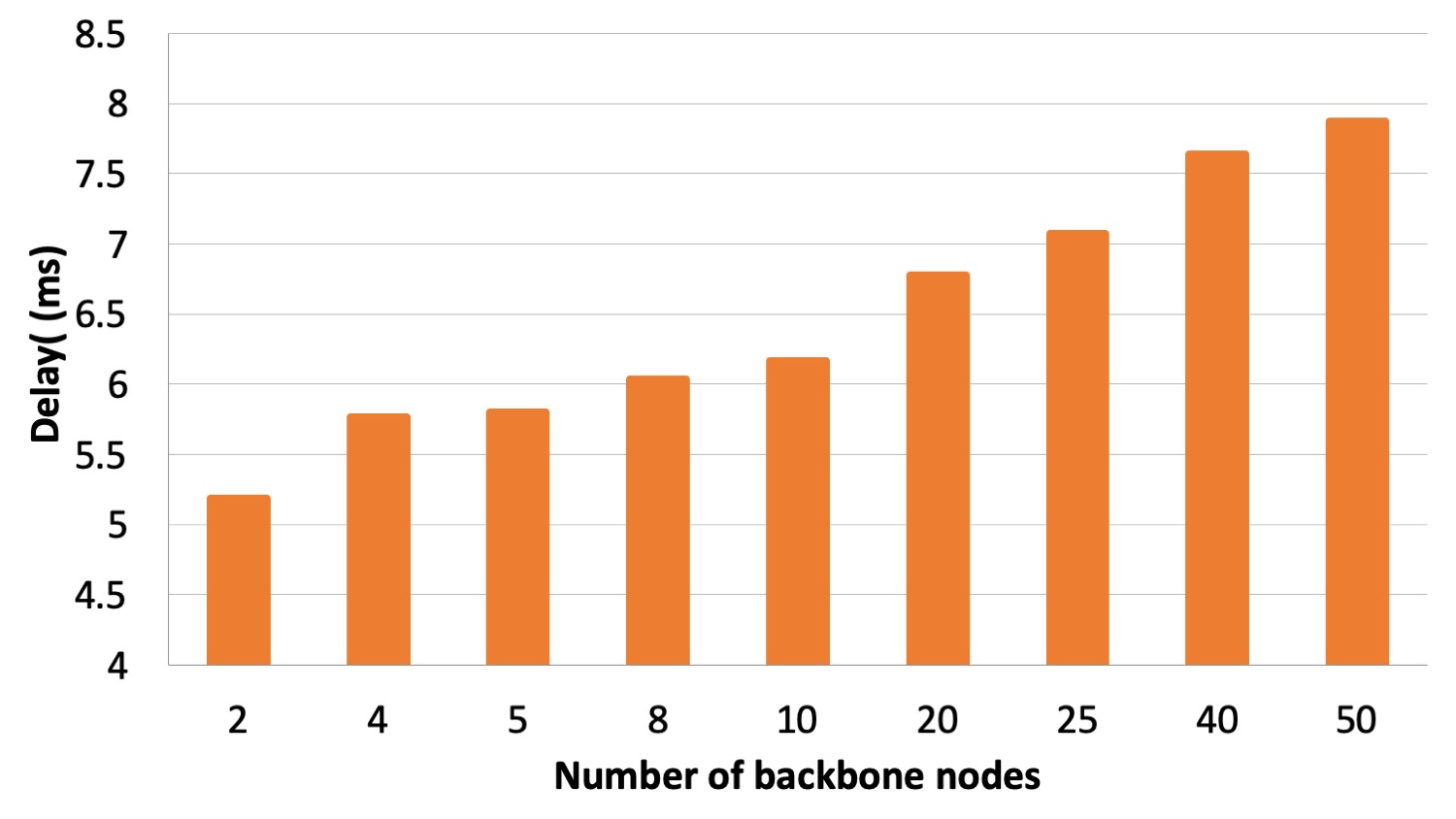}   
		\caption{Studying the communication delay while varying the number of backbone nodes. }
		\label{fig:ev:delaybb}
	\end{center}
\end{figure}

Recall from Table \ref{tab:scaling-performance} that  the delay in Vericom largely depends on the number of hops in a communication, thus we next study the impact of varying the number of backbone nodes  on the communication delay.  To prevent the traffic from traveling the same route while the number of backbone nodes increases, each new backbone node joins a randomly chosen backbone node and  the source and destination of the traffic are  randomly selected. Figure \ref{fig:ev:delaybb} outlines the simulation results. A 25-fold increase in the number of backbone nodes from 2 to 50 nodes results in a 50\% increase in communication delay.  \par 

\begin{figure}
	\begin{center}
		\includegraphics[width=8cm ,height=8cm ,keepaspectratio]{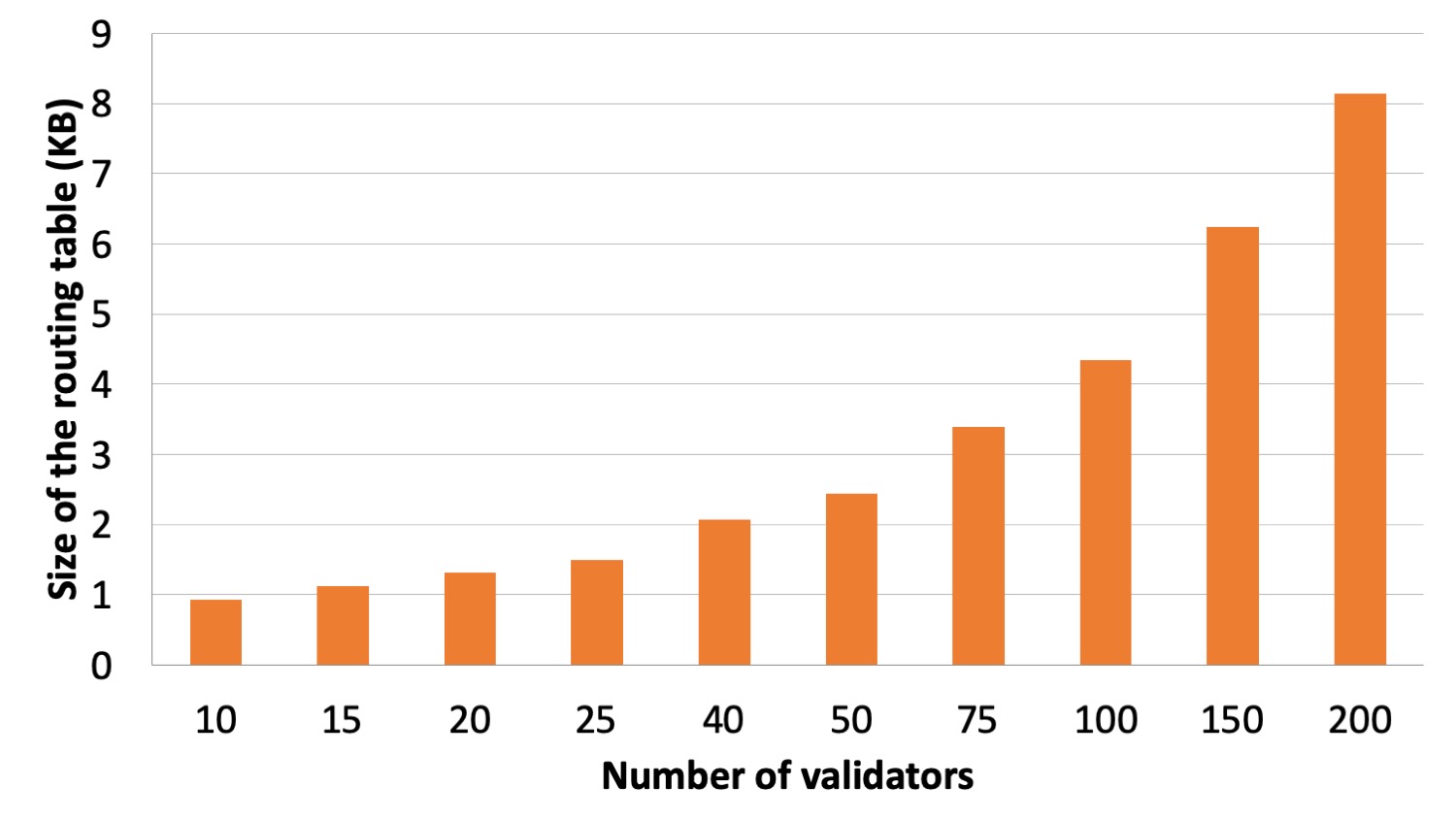}   
		\caption{Studying the size of the routing table in backbone nodes. }
		\label{fig:ev:routing-table}
	\end{center}
\end{figure}

\textit{Vericom overhead:} Finally we study the overheads incurred by Vericom. We first study the size of the routing table. As evident from the simulation results shown in Figure \ref{fig:ev:routing-table}, the size of the routing table increases from 0.9 KB to 8 KB while increasing the number of blockchain validators from 10 to 200. Recall from Section \ref{sec:architecture} that the routing table includes the PK of the validators and the IP address of the next hop backbone node for routing traffic. We next study the increased blocksize in Vericom to include the signature and PK of the verifiers. The size of the PK and signature in a transaction in our simulation is 459B. Thus, generally the  incurred overhead can be measured as (459 * \textit{V})B, where \textit{V} is the size of the verifier set. \par

\section{Conclusion }\label{sec:conclusion}
In this paper we introduced a Verification and Communication architecture for IoT-based blockchain known as Vericom. Vericom introduces a distributed yet scalable blockchain architecture by introducing a PK-based traffic, i.e., transactions and blocks, multicasting algorithm which in turn reduces the bandwidth consumption of the underlying IoT nodes as compared with conventional blockchains where traffic is broadcasted to all nodes.  Vericom incorporates two layers which are: i) transmission layer where a backbone network is introduces to route traffic,  and ii) verification layer where traffic is verified by a randomly selected set of nodes that are unique for each transaction or block which in turn   reduces the computational overhead associated with verifying new blocks and transactions. The simulation results prove that Vericom significantly reduces the blockchain packet and computational overhead which in turn facilitates the adaptation  of blockchain for low resource available IoT devices.    \par

\bibliography{mybibfile}

\end{document}